# Quantization, Doppler shift and the invariance of the speed of light via the invariance of the counted numbers of photons: An interesting pedagogical problem

## Bernhard Rothenstein[1] and Stefan Popescu[2]

1) Politehnica University of Timisoara, Physics Department,
Timisoara, Romania brothenstein@gmail.com
2) Siemens AG, Erlangen, Germany stefan.popescu@siemens.com

**Abstract**. We show that when the observers are located in a plane electromagnetic wave it isn't compulsory for them to take into account the time dilation and length contraction effects when the wave is detected from two inertial reference frames in relative motion. We also illustrate the difference between the approach that uses Einstein type of observers confronted with time dilation and length contraction effects and an unconventional approach that uses observers detecting information from the energy that arrives at their location.

## 1. Introduction

Margaritondo[1] presents a different way to introduce the concept of event and to derive the transformation equations for its energy. The basic idea considers a predetermined volume located inside a plane and linearly polarized electromagnetic wave. This volume includes a given number of photons and a given energy.

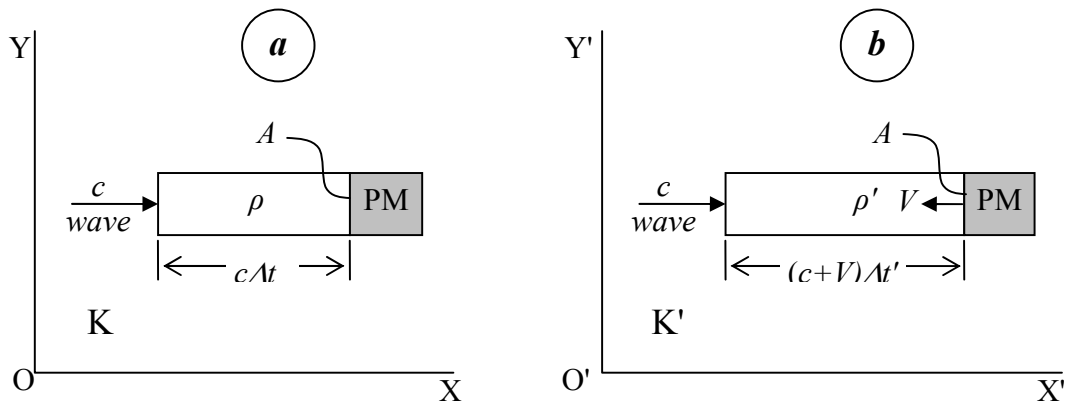

*Figure 1*. Scenario for deriving the transformation equation for the energy of a photon. Figure 1a depicts the situation detected from the rest frame of the photomultiplier whereas Figure 1b illustrates the situation detected from a reference frame which approaches the first one.

In order to make the problem transparent we considered the scenario presented in Figure 1, involving the mentioned electromagnetic wave propagating in empty space, incident on a photomultiplier (PM) whose ticks account for the incidence rate of photons.[2] First we consider the problem from the rest frame of the photomultiplier (K) and second from a reference frame K' being in the standard arrangement, with frame K moving with constant velocity $-V$ in the negative direction of the overlapped axes OX(O'X'). We consider the situation when the front of the electromagnetic wave arrives at the sensitive surface of the photomultiplier and so the wave interaction with the photomultiplier can be neglected. The wave energy stored into the photomultiplier is characterized by the *energy density* - $\rho$ in K and $\rho'$ in K' related in the case of this scenario by[3]

$$\rho' = \frac{1-\frac{V}{c}}{1+\frac{V}{c}} \rho \qquad (1)$$

and so the energy $W$ incident on the photomultiplier during a time interval $\Delta t$ (proper time interval) can be expressed as

$$W = \rho c A \Delta t \qquad (2)$$

where $cA\Delta t$ represents the volume in which the energy is stored, $A$ representing the active surface of the photomultiplier and $c$ the speed at which the wave propagates. The dimensions defining the surface $A$ are relativistic invariant being perpendicular to the direction of relative motion and therefore $A$ is also relativistic invariant.

When detected from K' the electromagnetic wave propagates with speed $c+V$ and we express the incident energy as

$$W' = \rho' A(c+V) \Delta t' \qquad (3)$$

where $\Delta t'$ represents a time interval related to the proper time interval by

$$\Delta t' = \frac{\Delta t}{\sqrt{1-\frac{V^2}{c^2}}} \qquad (4)$$

Let $Q$ and $Q'$ be the energy of the same photon detected from K and K' respectively. The counted number of stable particles being a relativistic invariant we have obviously $N=N'$. Further in K frame we have

$$N = \frac{\rho c A \Delta t}{Q} \qquad (5)$$

whereas in K' we have

$$N = \frac{\rho' A (c+V) \Delta t'}{Q'} . \qquad (6)$$

Eliminating $N$ between (5) and (6) we obtain the Lorentz transformation for the photon energy

$$Q' = Q \sqrt{\frac{1-\beta}{1+\beta}} \quad \text{with} \quad \beta = V/c \qquad (7)$$

Compared with the formula which accounts for the Doppler Effect relating the frequencies of the electromagnetic oscillations in the two frames ($v$ in K and $v'$ in K')[2]

$$v' = v\sqrt{\frac{1-\beta}{1+\beta}} \qquad (8)$$

we obtain that $Q=hv$ and $Q'=hv'$, with $h$ representing an universal invariant constant.

**2. A second possible approach: Working inside the electromagnetic wave**

Blatter and Greber[4] make a net distinction between the Einstein type of observers, which are equipped with meter sticks and synchronized clocks and are confronted with the time dilation and length contraction effects, and the observers collecting information about the events by using the light signals arriving at their location (photographic detection) or light signals they emit and receive back (radar detection). As we see, the scenario that we presented above mixes up physical quantities measured by the two types of observers. Besides, equations (5) and (6) are not in accordance with the first relativistic postulate that requires that the laws of physics should be the same in all inertial reference frames in relative motion and so should be the speed of light in empty space. We consider now that the observers of the two frames are entirely immersed in the electromagnetic wave. Observers from K separate a given volume $O = c \cdot A\Delta t$ in which the energy $W$ is stored and express the incident energy on the photomultiplier as

$$W = \frac{W}{A\lambda} cA\Delta t \qquad (9)$$

whereas observers from K' express the same energy as

$$W' = \frac{W'}{A\lambda'} cA\Delta t' . \qquad (10)$$

considering that the volume is $O' = c \cdot A\Delta t'$.

The electromagnetic wave offers its wavelength ($\lambda$ and $\lambda'$) as a particular unit to express the lengths along the OX and O'X' axes and the period of the electromagnetic oscillations ($\Delta t = T$, $\Delta t' = T'$) as another particular unit to express the time intervals. Further we also express the volumes as $A\lambda$ and $A\lambda'$ in K and in K' respectively. Taking into account the invariance of the counted number of photons we obtain from (9)

$$N = \frac{W}{Q} \qquad (11)$$

whereas from (10) we obtain

$$N = \frac{W'}{Q'} \qquad (12)$$

the energy transforming as

$$W' = W\sqrt{\frac{1-\beta}{1+\beta}} \qquad (13)$$

We obtain that the energy of the photon transforms in accordance with (7) recovering Margaritondo's results.

In order to complete the derivations, the volume $O$ in K and $O'$ in K' transforms as the wavelength does, i.e.

$$\frac{O'}{O} = \frac{\lambda'}{\lambda} = \frac{\nu}{\nu'} = \sqrt{\frac{1+\beta}{1-\beta}} \qquad (14)$$

and not in accordance with the length contraction effect.

### 3. Conclusions

Two approaches to the same scenario show the difference between the ways in which the two types of relativistic observers solve the same problem. The moral is that when working in the electromagnetic wave we shouldn't mix physical quantities measured by different kind of observers. We avoid the time dilation and length contraction effects by employing relativistic observers of second type and forcing them to use the wavelength and the wave period as length unit and respectively time unit instead of the meter stick and synchronized clock units as usual for the Einstein type of relativistic observers.

**References**
[1] G.Margaritondo, "Quantization, Doppler shift and invariance of the speed of light: some didactic problems and opportunities", Eur.J.Phys. **16** 169-171 (1996)
[2] P.Kocyk, P.Wiewior and C.Radzewicz, "Photon counting statistics: Undergraduate experiment", Am.J.Phys.64, 240 (1996)
[3] Richard Schlegel, "Radiation pressure on a rapidly moving surface", Am.J.Phys. 28, 687-694 (1960)
[4] H.Blatter and T.Greber, "Aberration and Doppler shift: An uncommon way to relativity", Am.J.Phys. 58, 942-945 (1990)